\documentclass[conference]{IEEEtran}
\IEEEoverridecommandlockouts
\usepackage{cite}
\usepackage{amsmath,amssymb,amsfonts}
\usepackage{algorithmic}
\usepackage{graphicx}
\usepackage{subcaption}
\usepackage{textcomp}
\usepackage{xcolor}
\usepackage{url}
\usepackage{stfloats}
\usepackage{mathtools}
\usepackage{float}
\usepackage{booktabs}
\usepackage[export]{adjustbox}
\def\BibTeX{{\rm B\kern-.05em{\sc i\kern-.025em b}\kern-.08em
    T\kern-.1667em\lower.7ex\hbox{E}\kern-.125emX}}

\usepackage[compact]{titlesec}         
\titlespacing{\section}{2pt}{2pt}{1pt} 
\AtBeginDocument{
  \setlength\abovedisplayskip{2pt}
  \setlength\belowdisplayskip{2pt}}

\begin{document}

\pagenumbering{gobble}

\title{Performance Analysis of Centralized and Distributed Massive MIMO for MTC}

\author{Eduardo Noboro Tominaga, Onel Luiz Alcaraz López, Hirley Alves, Richard Demo Souza\IEEEauthorrefmark{1}, Leonardo Terças\\

	\IEEEauthorblockA{
		\IEEEauthorrefmark{0}6G Flagship, Centre for Wireless Communications (CWC), University of Oulu, Finland\\
		\{eduardo.noborotominaga, onel.alcarazlopez, leonardo.tercas, hirley.alves\}@oulu.fi\\
		\IEEEauthorrefmark{1}Federal University of Santa Catarina (UFSC), Florian\'{o}polis, Brazil, richard.demo@ufsc.br\\
	}
}

\maketitle

\begin{abstract}
Massive Multiple-Input Multiple-Output (mMIMO) is one of the essential technologies introduced by the Fifth Generation (5G) of wireless communication systems. However, although mMIMO provides many benefits for wireless communications, it cannot ensure uniform wireless coverage and suffers from inter-cell interference inherent to the traditional cellular network paradigm. Therefore, industry and academia are working on the evolution from conventional Centralized mMIMO (CmMIMO) to Distributed mMIMO (DmMIMO) architectures for the Sixth Generation (6G) of wireless networks. Under this new paradigm, several Access Points (APs) are distributed in the coverage area, and all jointly cooperate to serve the active devices. Aiming at Machine-Type Communication (MTC) use cases, we compare the performance of CmMIMO and different DmMIMO deployments in an indoor industrial scenario considering regular and alarm traffic patterns for MTC. Our simulation results show that DmMIMO's performance is often superior to CmMIMO. However, the traditional CmMIMO can outperform DmMIMO when the devices' channels are highly correlated.
\end{abstract}


\section{Introduction}


\par The Fifth Generation (5G) of wireless networks is already a reality. However, current 5G technologies focus on providing very high spectral efficiencies for enhanced Mobile BroadBand (eMBB) use cases, and are not yet able to fully meet the demanding requirements of critical Machine-Type Communications (cMTC) and massive MTC (mMTC) use cases \cite{6G_MTC_White_Paper}. 

\par One of the major physical layer enhancements introduced in 5G is massive Multiple-Input Multiple-Output (mMIMO) \cite{bjornson2017}, which consists of the use of Base Stations (BSs) with a very large number of antenna elements. mMIMO provides several benefits for all 5G use cases: very high beamforming gains (thus making it possible to achieve very high spectral efficiencies), high spatial multiplexing capabilities (potentially servicing multiple devices simultaneously), and quasi-deterministic wireless links (important for applications with stringent requirements in terms of latency and reliability). However, owing to the traditional cellular network paradigm, mMIMO cannot provide uniform wireless coverage. That is, users at the cell boundaries have poor signal coverage compared to users closer to the BS. Thus, 
several recent works have been proposing the evolution from Centralized mMIMO (CmMIMO) architectures to Distributed mMIMO (DmMIMO) architectures, also known as Cell-Free (CF) mMIMO, for beyond-5G and 6G networks. According to this new paradigm, there are no longer cell boundaries: several Access Points (APs) are distributed in the coverage area, each one of them connected to a common Central Processing Unit (CPU) through a fronthaul connection. Then, all the APs jointly cooperate to serve all the devices in the network. Such approach provides a more uniform wireless coverage to all users in the network \cite{demir2021}. 


\par Notably, most of the recent works about DmMIMO (e.g., \cite{ngo2017,emil2020,interdonato2019}) investigate the performance benefits of this novel paradigm in terms of spectral efficiency by focusing on eMBB use cases. Only a few papers study the benefits of distributed mMIMO for cMTC  (e.g. \cite{casciano2019,nasir2021}) and mMTC (e.g. \cite{ke2021,ganesan2021}). Moreover, different works consider different spatial distributions for the APs, e.g., arbitrarily distributed on the coverage area \cite{ngo2017,emil2020}, uniform deployment of APs on a ``grid" \cite{interdonato2019} or around the borders of the coverage area on a ``linear"\footnote{The linear deployment is inspired in the Radio Stripes developed by Ericsson in 2017\cite{interdonato2019}. Aiming at a DmMIMO network that requires only one fronthaul connection to the CPU, a single radio stripe is deployed across the borders of the coverage area, and sequential processing is utilized \cite{shaik2020}.} fashion \cite{shaik2020}.


\par In our previous work \cite{tominaga2022}, we compared the performance of CmMIMO and DmMIMO in terms of received signal strength and variability at a single active device. We compared grid and linear deployments of APs considering the cases of single-antenna APs and multi-antenna APs. The performance of a grid deployment and a linear deployment of APs in the downlink of an indoor industrial scenario was also compared in \cite{fernandes2022}, where the authors investigated the effects of isolated and cumulative failures on the hardware of APs. They also proposed protection schemes that strongly or entirely mitigate the effects of these failures.


\par In this paper, we extend the framework of our previous work by comparing the legacy CmMIMO architecture with DmMIMO architectutes with grid and linear deployments of APs \cite{tominaga2022} in a multi-user scenario. While several works such as \cite{ngo2017,emil2020,interdonato2019} evaluated the spectral efficiency aspects of DmMIMO, which is a performance metric of interest for eMBB, here we evaluate its reliability and massive connectivity aspects. Focusing on indoor industrial applications using a wireless channel model validated by 3GPP for such scenarios \cite{Path_Loss_Models_Industrial}, we aim at evaluating the impact of different traffic models on the performance of CmMIMO and DmMIMO. 
Note that while related works dealing with the performance of DmMIMO in MTC networks considered only arbitrarily distributed Machine-Type Devices (MTDs) \cite{ke2021},\cite{ganesan2021}, here we study the performance of DmMIMO under two different traffic models: i) regular traffic, for which we consider that the active MTDs are uniformly distributed in the coverage area; and ii) alarm traffic, for which the active MTDs are concentrated nearby the epicenter of alarm events, i.e., they present a high spatio-temporal correlation. 
Moreover, while works dealing with traffic models for MTC have obtained closed-form expressions only for temporal-correlation of the MTC traffic under alarm events \cite{laner2013,katerina2014,thomsen2017}, we propose a closed-from expression for the spatial distribution of active MTDs under alarm traffic. We reveal that, under regular traffic, the grid distribution of APs always presents the better performance compared to the linear deployment and to CmMIMO. Here, the performance of the linear deployment is only slightly better than the performance of CmMIMO. But interestingly, in the case of alarm traffic, there are situations where CmMIMO outperforms DmMIMO.

\begin{figure*}[b]
    $$\mathclap{\rule{18cm}{0.5pt}}$$
    \begin{equation}
        \tag{3}
        \label{positionAPsRadioStripe}
        (x_{q},y_{q},z_{q})=
        \begin{cases}
            \vspace{2pt}
            \left[\left(q_B-\dfrac{1}{2} \right)\dfrac{4l}{Q},0,h\right],\;q_B\in\{1,2,\ldots,Q/4\},\text{ if the AP is on the bottom wall;}\\
            \vspace{2pt}
            \left[0,\left(q_L-\dfrac{1}{2} \right)\dfrac{4l}{Q},h\right],\;q_L\in\{1,2,\ldots,Q/4\},\text{ if the AP is on the left wall;}\\
            \vspace{2pt}
            \left[\left(q_T-\dfrac{1}{2} \right)\dfrac{4l}{Q},l,h\right],\;q_T\in\{1,2,\ldots,Q/4\},\text{ if the AP is on the top wall;}\\
            \vspace{2pt}
            \left[l,\left(q_R-\dfrac{1}{2} \right)\dfrac{4l}{Q},h\right],\;q_R\in\{1,2,\ldots,Q/4\},\text{ if the AP is on the right wall.}\\
        \end{cases}
    \end{equation}    
\end{figure*}

\vspace{-0.07cm}
\par This paper is organized as follows: Section \ref{System Model} comprises the system model: the different mMIMO deployment schemes, the signal model, and the wireless channel model for indoor industrial scenarios. Traffic models for MTC are discussed in Section \ref{Traffic Models}. Monte Carlo simulation results comparing the performance of the different mMIMO deployment schemes are presented in Section \ref{Numerical Results}. Finally, the conclusions of this work are drawn in Section \ref{Conclusions} (\textit{Reproducible Research:} simulation codes utilized for this paper will be available in https://github.com/eduardotominaga).

\par \textbf{Notation:} lowercase bold face letters denote column vectors, while boldface upper case letters denote matrices. $a_i$ is the $i$-th element of the column vector $\textbf{a}$, while $\textbf{a}_i$ is the $i$-th column of the matrix $\textbf{A}$. $A_{i,j}$ is the $i$-th row, $j$-th column element of the matrix $\textbf{A}$. $\textbf{I}_M$ is the identity matrix with size $M\times M$. The superscripts $(\cdot)^T$ and $(\cdot)^H$ denote the transpose and the conjugate transpose operation, respectively. The magnitude of a scalar quantity or the cardinality of a set is denoted by $|\cdot|$. The Euclidian norm is denoted by $\Vert\cdot\rVert$. We denote the circularly symmetric complex Gaussian distribution with mean $\mathbf{a}$ and covariance $\mathbf{B}$ by $\mathcal{CN}(\mathbf{a},\mathbf{B})$.

\section{System Model}
\label{System Model}



\par We consider the uplink of an indoor industrial scenario with dimensions $l\times l$, which represents a factory hall. $K_{\text{total}}$ single-antenna MTDs are jointly served by a total number of $Q$ APs, each equipped with $S=M/Q$ antenna elements, where $M$ is the total number of antenna elements. All MTDs communicate to all the APs in the same time-frequency resource. Our analysis is applicable to either a narrowband single-carrier system or a narrowband subcarrier within a multi-carrier system.

\par
\begin{figure}[H]
    \centering
    \includegraphics[scale=0.35]{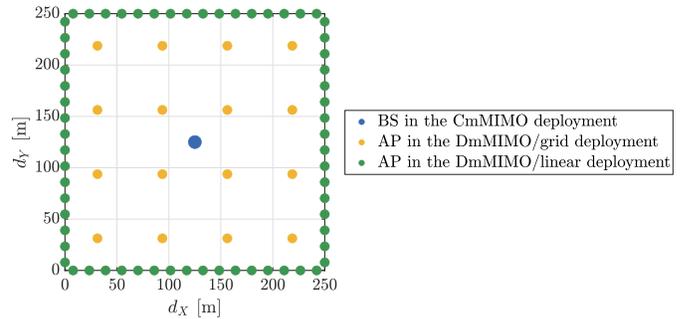}
    \caption{Illustration of the different mMIMO deployments considered in this work given $l=250$ m. In the CmMIMO deployment, a single BS is located at the center of the square area. In the DmMIMO with grid deployment, the APs are uniformly distributed in the square area. Finally, in the DmMIMO with linear deployment, the APs are serially connected around the square area.}
    \label{deployments}
\end{figure}

 Denote $K$ as the number of active MTDs in a given time slot, and assume that the total number of antenna elements is at least equal to the number of active MTDs, i.e., $M\geq K$. The $Q$ APs can be deployed in the factory hall according to one of the mMIMO deployment schemes discussed in the following and illustrated in Fig. \ref{deployments}.


\subsection{mMIMO Architectures}
\label{mMIMO_architectures}

\subsubsection{CmMIMO}
    
\par A single BS with $M$ antenna elements is located at the position
\begin{equation}
    \tag{1}
    (x_{\text{BS}},y_{\text{BS}},z_{\text{BS}})=(l/2,l/2,h),
\end{equation}
that is, the center of the factory hall at height $h$. 

\subsubsection{DmMIMO, grid deployment}
    
\par $Q$ APs are uniformly distributed on the ceiling of the factory hall. Each AP has $S=M/Q$ antenna elements, selected such that $S \in \mathbb{Z}$.  Let $(q_x,q_y)$ denote the indexes of the $q$-th AP in the grid of APs, where $q_i=\{1,2,\ldots,\sqrt{Q}\},\;i\in\{x,y\}$. The coordinates of the $q$-th AP in the square coverage area are then given by
\begin{equation}
    \tag{2}
    (x_{q},y_{q},z_{q})=\!\left[\!\!\left(\!q_x-\dfrac{1}{2}\!\right)\!\dfrac{l}{\sqrt{Q}},\!\left(\!q_y-\dfrac{1}{2}\!\right)\!\dfrac{l}{\sqrt{Q}},h\!\right]\!,
\end{equation}
where $h$ is the height of the AP in the factory hall.

\subsubsection{DmMIMO, linear deployment}

\par The APs are serially connected around the factory hall. The coordinates of the $q$-th AP are given by (\ref{positionAPsRadioStripe}).

\subsection{Signal Model}
\label{Signal Model}

\setcounter{equation}{3}

\par The $M\times 1$ collective vector of received samples is
\begin{equation}
    \textbf{y}=\sqrt{p_u}\textbf{G}\textbf{x}+\textbf{n},
\end{equation}
where $p_u$ is the fixed uplink transmit power common to all MTDs, $\textbf{G}\in\mathbb{C}^{M\times K}$ is the channel matrix between the $M$ antenna elements and the $K$ active MTDs, $\textbf{x}~\sim~\mathcal{CN}(\mathbf{0}_{M\times 1},\mathbf{I}_K)$ is the vector of symbols simultaneously transmitted by the $K$ MTDs, and $\textbf{n}\sim \mathcal{CN}(\mathbf{0}_{M\times 1},\sigma^2\mathbf{I}_M)$ is the vector of Additive White Gaussian Noise (AWGN) samples. The noise power (in Watts) is given by
\begin{equation}
    \sigma^2=N_0BN_F,
\end{equation}
where $N_0=10^{-14.4}\text{ W/Hz}$ is the Power Spectral Density (PSD) of the thermal noise, $B$ is the bandwidth in Hz, and $N_F$ is the noise figure at the receivers.

\par The collective vector of the wireless channel coefficients between the $k$-th MTD and the $Q$ APs is $\textbf{g}_k=[\textbf{g}_{k,1}^T,\textbf{g}_{k,2}^T,\ldots,\textbf{g}_{k,Q}^T]^T \in \mathbb{C}^{M\times1}$, where $\textbf{g}_{k,q} = \sqrt{\beta_{kq}}\textbf{h}_{k,q}^T\in \mathbb{C}^{S\times1}$ is the vector of wireless channel coefficients between the $k$-th MTD and the $q$-th AP, $\beta_{kq}$ is the large scale fading coefficient between the $k$-th MTD and the $q$-th AP, and $\textbf{h}_{k,q}\in\mathbb{C}^{S\times1}$ is the vector of small scale fading coefficients from the $k$-th MTD to the $q$-th AP. 
The matrix $\textbf{G}\in\mathbb{C}^{M\times K}$ containing the channel vectors of the $K$ MTDs can be written as
\begin{equation}
    \textbf{G}=[\textbf{g}_1,\textbf{g}_2,\ldots,\textbf{g}_K].
\end{equation}

\par We assume that the CPU has perfect Channel State Information (CSI). Let $\textbf{V}\in\mathbb{C}^{M\times K}$ be a linear detector matrix used for the joint decoding of the signals transmitted from the $K$ MTDs at all the APs. The received signal after the linear detection operation is split in $K$ streams and given by
\begin{equation}
    \textbf{r}=\textbf{V}^H\textbf{y}=\sqrt{p_u}\textbf{V}^H\textbf{G}\textbf{x}+\textbf{V}^H\textbf{n}.
\end{equation}

\par Let $r_k$ and $x_k$ denote the $k$-th elements of $\textbf{r}$ and $\textbf{x}$, respectively. Then, the received signal corresponding to the $k$-th MTD can be written as
\begin{equation}
    \label{r_k}
    r_k=\sqrt{p_u}\textbf{v}_k^H\textbf{g}_kx_k + \sqrt{p_u}\textbf{v}_k^H\sum_{k'\neq k}^K \textbf{g}_{k'}x_{k'} + \textbf{v}_k^H\textbf{n},
\end{equation}
where $\textbf{v}_k$ and $\textbf{g}_k$ are the $k$-th columns of the matrices $\textbf{V}$ and $\textbf{G}$, respectively. The first term in (\ref{r_k}) corresponds to the signal of interest, while the remaining terms in the sequence correspond to the interference from other MTDs and noise.

\par From (\ref{r_k}), the Signal-to-Interference-plus-Noise Ratio (SINR) of the uplink transmission from the $k$-th MTD to all the APs is given by
\begin{equation}
    \label{gamma_k}
    \gamma_k=\dfrac{p_u|\textbf{v}_k^H\textbf{g}_k|^2}{p_u\sum_{k'\neq k}^K |\textbf{v}_k^H\textbf{g}_{k'}^2|+\sigma^2\lVert\textbf{v}_k^H\rVert^2}.
\end{equation}


\par In this work, we adopt the centralized Minimum Mean Square Error (MMSE) combining, which outperforms other linear schemes such as Maximum Ratio Combining (MRC) and Zero Forcing (ZF) \cite{tominaga2022}. The linear detection matrix for MMSE combining is \cite{liu2016}
\begin{equation}
    \textbf{V}=\left(\textbf{G}\textbf{G}^H+\dfrac{\sigma^2}{p_u}\textbf{I}_N\right)^{-1}\textbf{G}.
\end{equation}
The receive beamforming vector for the $k$-th user is the $k$-th column of the matrix $\textbf{V}$.


\par Let $D$ denote the number of correctly decoded MTDs in a given time slot. As the performance metric, we consider the outage probability, which is defined as
\begin{equation}
    \mathcal{P}_{\text{out}}=1-\dfrac{\mathbb{E}\left\{D\right\}}{K},
\end{equation}
where the data message of the $k$-th active MTD is assumed to be correctly decoded if $\gamma_k\ge 2^R-1$. Here, $R$ is the target data rate in bits/s/Hz. The outage probability is an important performance metric for MTC since it relates to the packet error rates and consequently to the number of packet retransmissions, i.e., the reliability and latency aspects of MTC networks.

\subsection{Wireless Channel Model}
\label{Wireless_Channel_Model}

\par We adopt a channel model validated by 3GPP for indoor industrial scenarios \cite{Path_Loss_Models_Industrial}. Despite its simplicity, this model was obtained through real measurements in a factory hall. The system operates with a carrier frequency $f_c=3.5$ GHz, since the 3.4 - 4.2 band is the main candidate for industrial private networks with bandwidths of up to 100 MHz \cite{Path_Loss_Models_Industrial}. We adopt the large scale fading model for indoor industrial scenarios from \cite{Path_Loss_Models_Industrial}, which assumes the existence of non line-of-sight between the MTDs and the APs. Moreover, the $S\times 1$ channel vector between the $k$-th MTD and the $q$-th AP is denoted by $\textbf{g}_{k,q}\sim\mathcal{CN}(\textbf{0}_{S\times 1},\beta_{kq}\textbf{I}_S)$, where $\beta_{kq}<1$ is the large scale fading term that accounts for path loss and shadowing.


\par The large-scale fading is modeled as
\begin{equation}
    \label{path loss}
    \overline{\text{PL}}_{kq}\text{[dB]}=32.5+20\log_{10}f_c+10\eta\log_{10}d_{kq},
\end{equation}
where $\eta=3.19$ is the path loss exponent in the considered scenario and $d_{kq}$ is the 3D distance between the $k$-th MTD and the $q$-th AP in meters. Meanwhile, the total attenuation due to distance and shadowing is
\begin{equation}
    \text{PL[dB]}_{kq} = \overline{\text{PL}}_{kq}\text{[dB]} + X_{\sigma_S}\text{[dB]},
\end{equation}
where $\overline{\text{PL}}_{kq}\text{[dB]}$ is given by (\ref{path loss}) and $X_{\sigma_S}\text{[dB]}\sim\mathcal{N}(0,\sigma_S^2)$ is a log-normal RV that represents the shadowing term, with $\sigma_S=7.56$ dB \cite{Path_Loss_Models_Industrial}. The large scale fading coefficient between the $k$-th MTC device and the $q$-th AP is
\begin{equation}
    \beta_{kq}=1/\text{PL}_{kq}(d_{kq}).
\end{equation}

\par
\begin{figure}[t]
    \centering
    \includegraphics[scale=0.6]{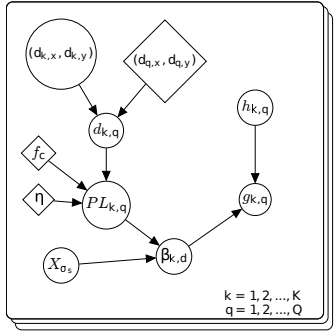}
    \caption{Probabilistic graphical model for the wireless channel vectors of the active MTDs.}
    \label{Graphical_Model}
\end{figure}

 Note that the wireless channel vectors depend on the positions of the active MTDs and APs, on the path losses owing to the propagation distances, on the shadowing, and also on the fast fading coefficients. In order to capture all these dependencies, we resort to Probabilistic Graphical Model (PGM) \cite{koller2009}, which is a tool used to represent the relationships among the variables and parameters via a graph. With the PGM at hand, we calculate the resulting joint probability distribution or other relevant statistic. For this particular case, the PGM is illustrated in Fig. \ref{Graphical_Model}, which yields a compact representation of such multidimensional model. Then, the probabilistic joint distribution is
\begin{equation}
    \begin{aligned}
    \centering
         f(V) = & \prod_{q=1}^Q \bigg[ f(x_{q},y_{q}) \prod_{k=1}^K \bigg[ f(d_{kq} | (x_{q},y_{q}), (x_{k},y_{k})) \\ & f(X_{\sigma_{s}}) f(\text{PL}_{kq} | f_c, \eta, d_{kq}) f(\beta_{kq}| \text{PL}_{kq}, X_{\sigma_s})f(h_{kq}) \\ & f(\textbf{g}_{kq} | \textbf{h}_{kq}, \beta_{kq}) \bigg] \bigg].          
    \end{aligned}
    \raisetag{1cm}
\end{equation}

\section{Traffic Models for MTC}
\label{Traffic Models}

\par The MTDs can operate in two different modes \cite{thomsen2017}: regular and alarm. In the case of regular mode, the MTDs send periodic updates containing information about their status and the physical quantities they are measuring. In this situation, the MTD traffic is uncorrelated in time and space. Let $(x_{k},y_{k},h_{\text{MTD}})$ denote the coordinates of the position of the $k$-th active MTD. Thus, in the case of regular traffic, we can model the spatial distribution of active MTDs in a given time slot using a uniform distribution: $x_{k},y_k\sim\mathcal{U}[0,l]$.


\par
\begin{figure}[t]
    \centering
    \includegraphics[scale=0.35]{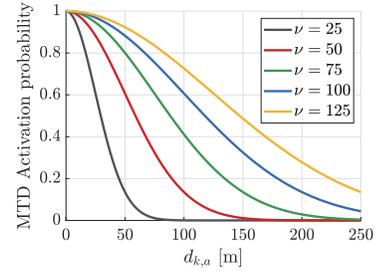}
    \caption{Probability of activation of a MTD versus the distance between the MTD and the epicenter of the alarm event, for different values of the intensity of the alarm event.}
    \label{figureATPF}
\end{figure}

 In a MTC network, there are often alarm events that trigger the activation of several devices located closely to the epicenter of the event. In other words, when such event occurs, the activation of the MTDs is highly correlated over the space and time. 
Let $A$ denote the number of alarm events occurring simultaneously in the factory hall, and  let $(x_{a},y_{a},z_{a})$ denote the coordinates of the epicenter of the $a$-th alarm event. When such an alarm event occurs, it triggers the activation of the MTDs located nearby. Thus, we model the probability of a MTD being activated owing to the occurrence of the alarm event using a scaled one-tailed Gaussian function as the Alarm Triggering Probability Function (ATPF), i.e.,\footnote{We chose this ATPF aiming at obtaining a closed-form expression for the PDF of spatial distribution of the active MTDs given the epicenter of an alarm event. Other options could be a linear function or an exponential function. In practice, an ATPF could be any positive and strictly decreasing function $f(d)$ defined for $d\geq0$ such that $f(0)=1$ and $\lim_{d\rightarrow\infty}f(d)=0$.}
\begin{equation}
    \label{equationATPF}
    \mathcal{P}_{\text{act}}(d_{ak},\sigma)=\exp\left(-\dfrac{d_{ak}^2}{2\nu^2}\right),
\end{equation}
where $d_{ak}$ is the distance (in meters) between the epicenter of the alarm event and the $k$-th MTD, and $\nu$ is a parameter that represents the intensity of the alarm event. In Fig. \ref{figureATPF}, we show the probability of activation of a MTD versus $d_{ak}$ for different values of $\nu$, considering the ATPF given by (\ref{equationATPF}).

\par Note that the shape of the ATPF in (\ref{equationATPF}) determines the spatial distribution of the active MTDs. Using this ATPF and assuming that the propagation of the alarm triggering occurs equally in both $x$-axis and $y$-axis, the location of the MTDs triggered by the alarm event follows a truncated\footnote{Note that we must guarantee the condition $x_{k},y_k\in[0,l],\;i\in\{x,y\}$ to ensure that the active MTDs are located inside the factory hall. This corresponds to the truncation of the 2D Gaussian PDF.} 2D Gaussian Probability Density Function (PDF) centered at $(x_{a},y_{a},z_{a})$:
\begin{equation}
    f(x,y|a) = \!\dfrac{1}{2\pi\nu^2}\! \!\left\{\!-\dfrac{1}{2}\!\left[\! \!\left(\!\dfrac{x-x_{a}}{\nu}\right)^2 \!+\! \!\left(\!\dfrac{y-y_{a}}{\nu}\right)^2 \!\right]\! \!\right\}\!,
\end{equation}   

\par
\begin{figure*}[t]
    \centering
    \begin{subfigure}[b]{0.3\textwidth}
        \centering
        \includegraphics[scale=0.35]{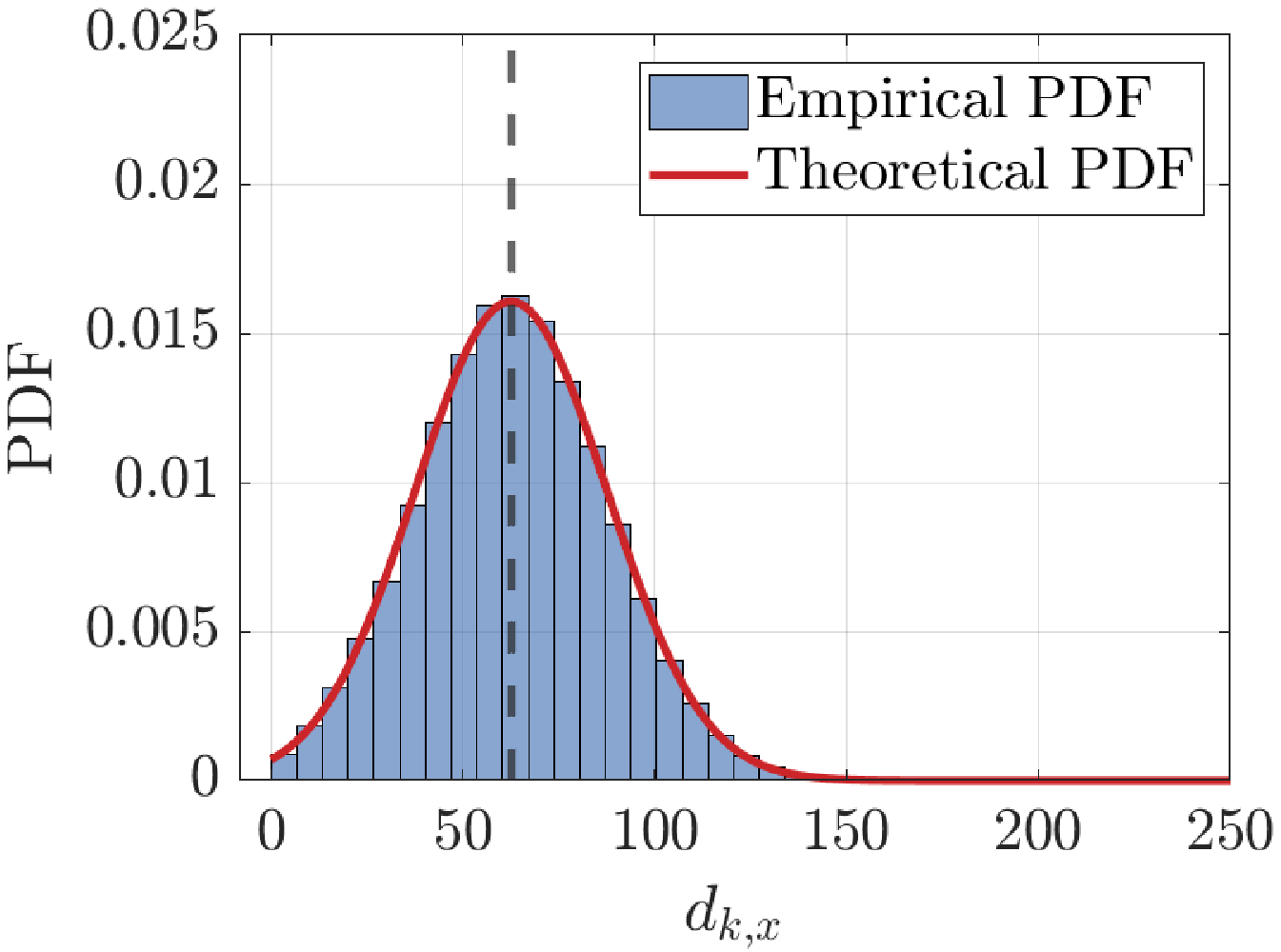}
        \caption{}
        \label{PDF_x}
    \end{subfigure}
    \hfill
    \begin{subfigure}[b]{0.3\textwidth}
        \centering
        \includegraphics[scale=0.35]{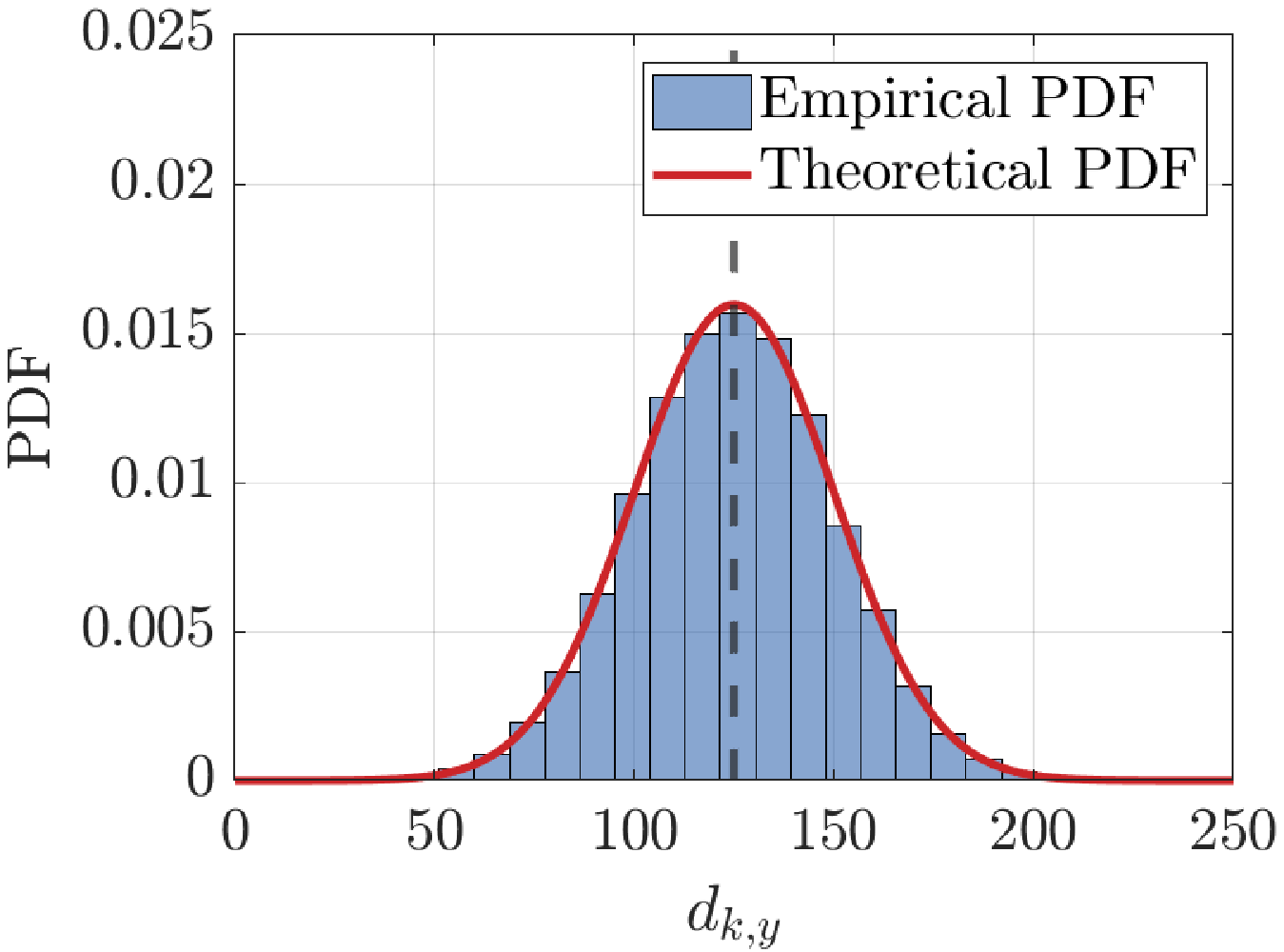}
        \caption{}
        \label{PDF_y}    
    \end{subfigure}
    \hfill
    \begin{subfigure}[b]{0.3\textwidth}
        \centering
        \includegraphics[scale=0.28]{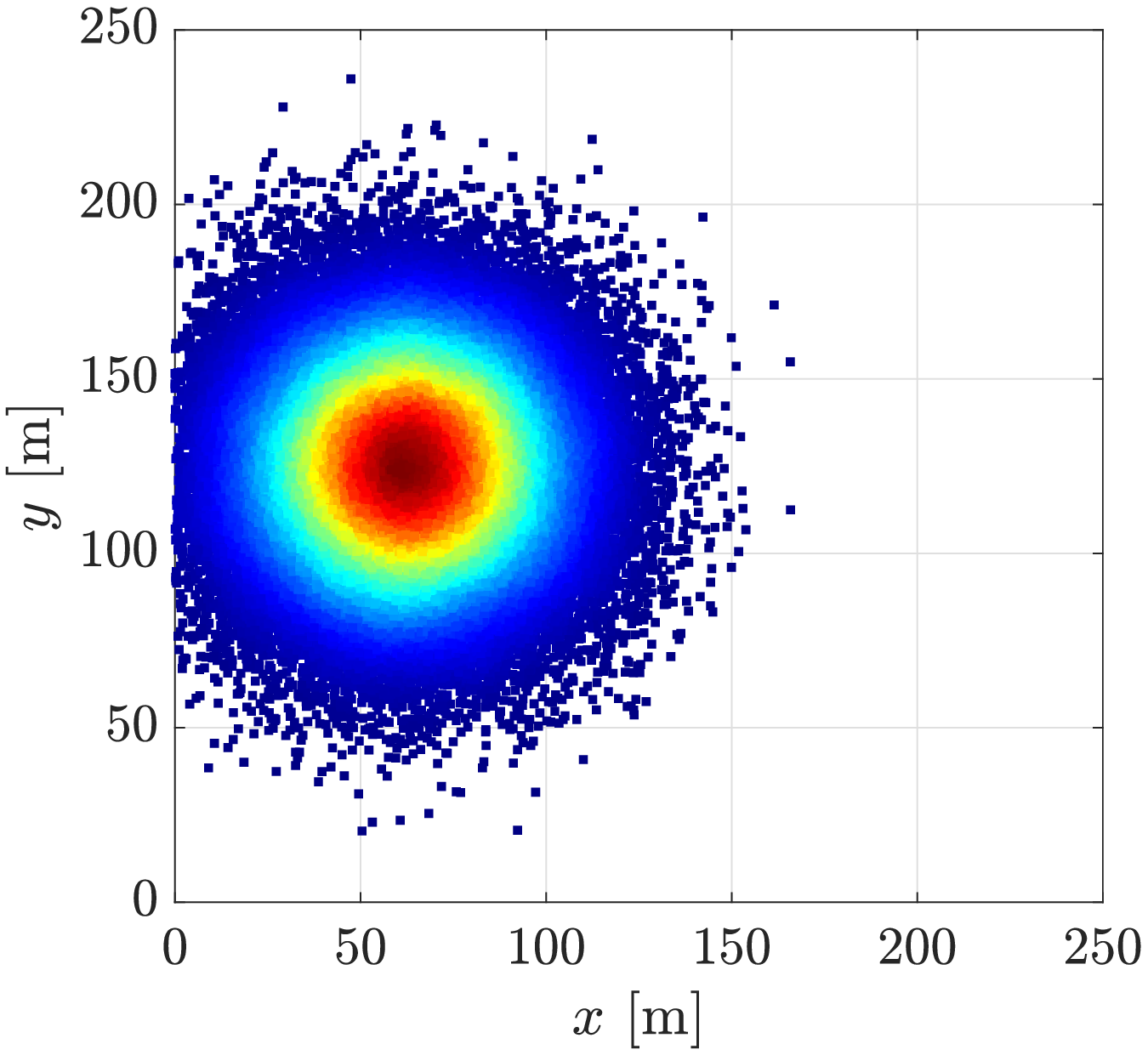}
        \caption{}
        \label{heatmapActiveMTDs}
    \end{subfigure}
    \caption{Empirical and theoretical PDFs of the location of active MTDs under the occurrence of an alarm event on the $x$-axis (a) and $y$-axis (b), and the heat map of the location of the triggered MTDs (c) for $l=250$ m, $(x_{a},y_{a},z_{a})=(l/4,l/2,0)$, $\nu=25$, $K_{\text{total}}=10^3$ and $N=10^3$ network realizations.}
    \label{distributionMTDs}
\end{figure*}

 In Figs. \ref{PDF_x} and \ref{PDF_y}, we show the empirical and theoretical PDFs of the location of active MTDs under an alarm event on the $x$-axis and $y$-axis, respectively. The heat map of the location of the triggered MTDs is shown in Fig.~\ref{heatmapActiveMTDs}. The parameters considered in Figs. \ref{distributionMTDs} are: $l=250$~m, $(x_{a},y_{a},z_{a})=(l/4,l/2,0)$, $\nu=25$, $K_{\text{Total}}=10^3$ and $N=10^3$ network realizations.

\section{Numerical Results}
\label{Numerical Results}

\par In this section, we resort to Monte Carlo simulations to compare the performance of the different mMIMO deployment schemes in the cases of regular traffic and alarm traffic. The simulation parameters are listed in Table \ref{tableSimulationParameters}. In the case of alarm traffic, we assume the occurrence of a single alarm event in the factory hall and, for the sake of simplicity, there is no regular traffic during the alarm event. 

\begin{figure}[t]

    \begin{subfigure}[t]{0.2\textwidth}
        \centering
        \includegraphics[scale=0.32]{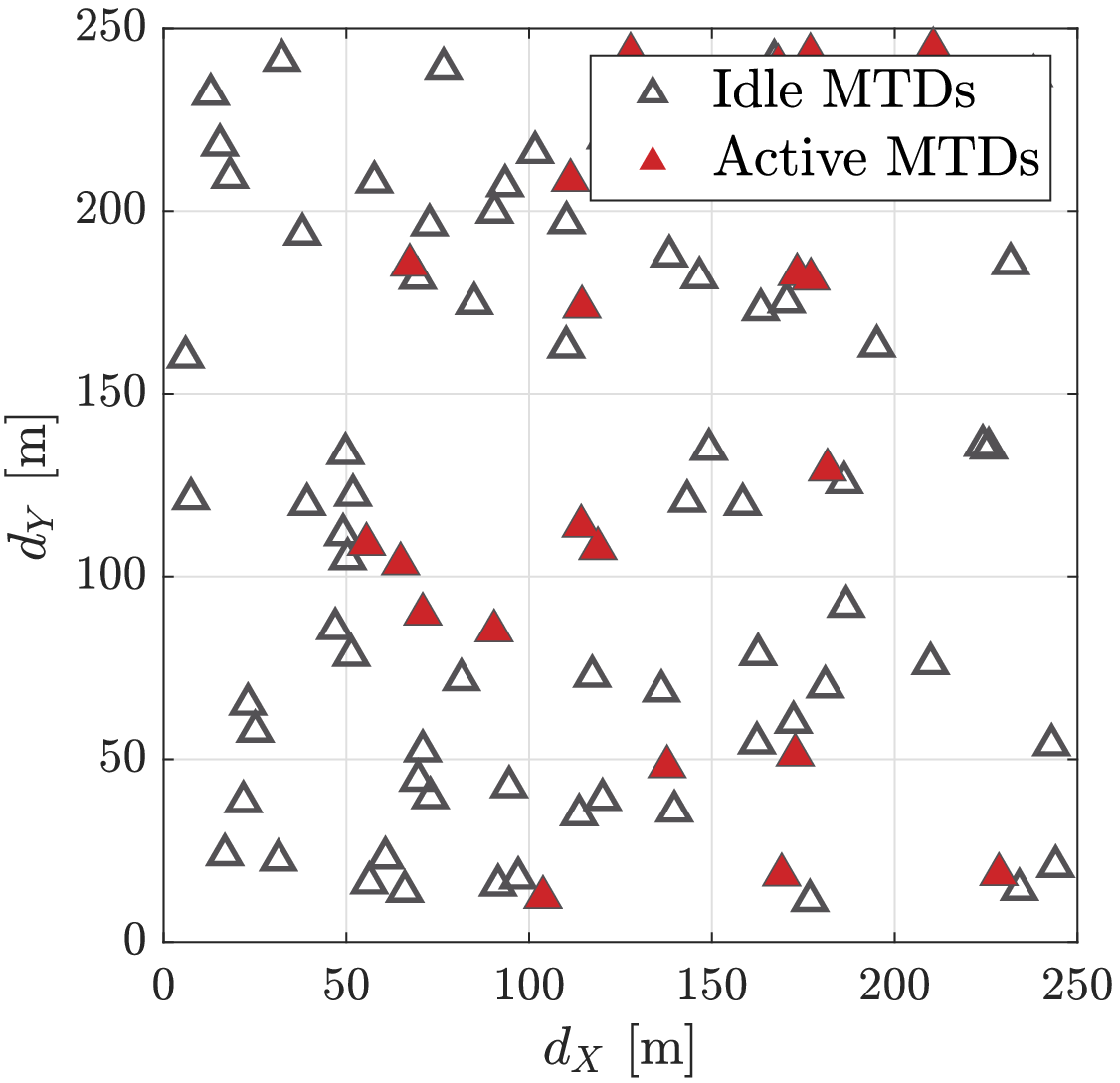}
        \caption{}
        \label{Regular Traffic}
    \end{subfigure}
    ~
    \begin{subfigure}[t]{0.2\textwidth}
        \centering
        \includegraphics[scale=0.32]{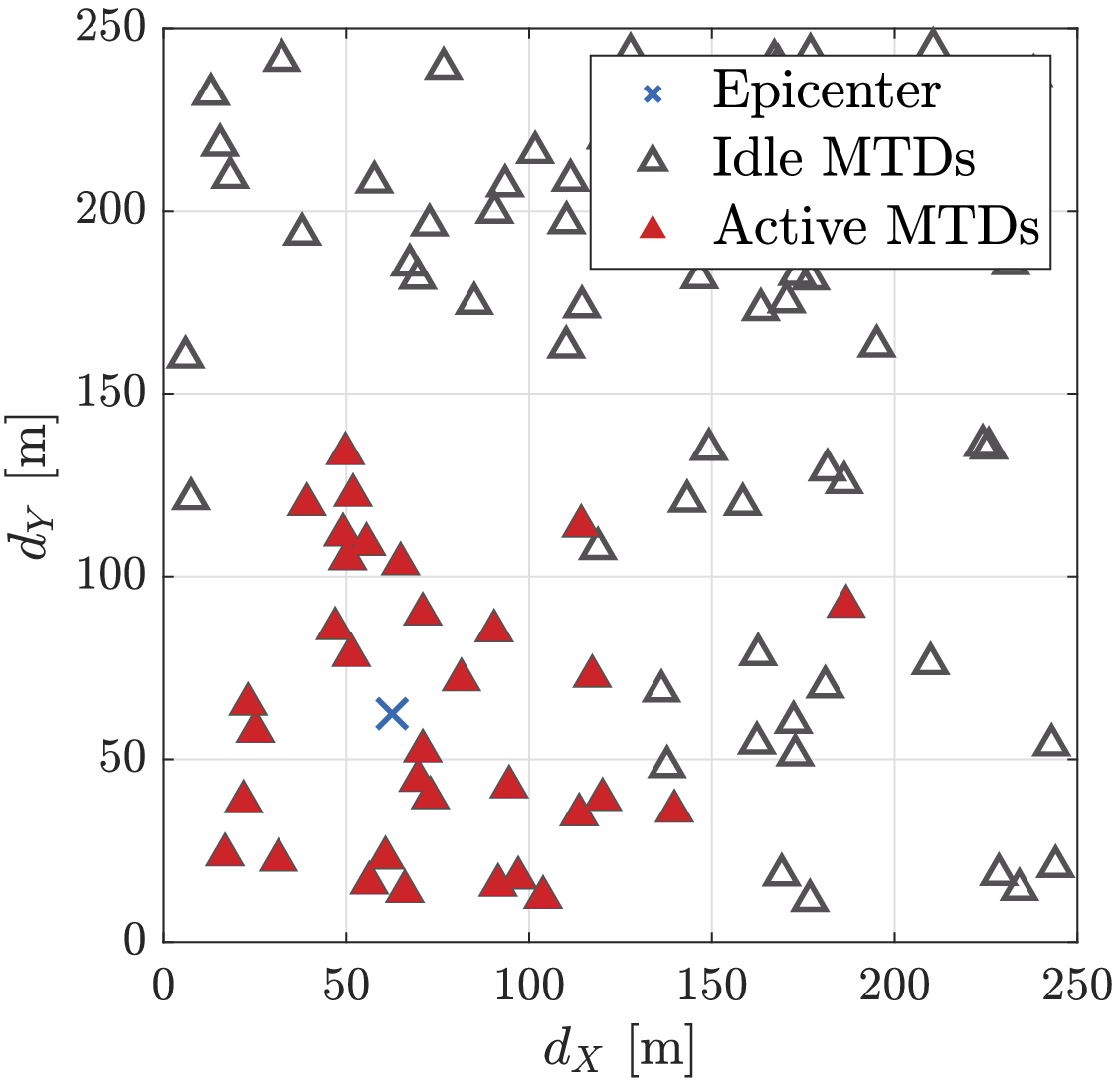}
        \caption{}
        \label{Alarm Traffic}    
    \end{subfigure}

    \caption{Snapshots of the indoor industrial network during regular traffic (a) and alarm traffic (b) for $l=250$ m.}
    \label{mMIMO Setups}
\end{figure}

\par We aim at computing the outage probability of the system. In order to achieve this goal, we generate $N=10^2$ different network realizations, i.e., different set of random positions for the $K$ active MTDs\footnote{In a real network, the number of active MTDs in each time slot is a RV. However, in order to compare the performance of the different settings, we set a number $K$ of active MTDs. In case of regular traffic, they are arbitrarily distributed, while they follow the PDF (\ref{equationATPF}) in case of alarm traffic.}. Then, for each network realization, we generate $10^3$ realizations for the matrix $\textbf{G}$. 

\begin{table}[]
    \centering
    \caption{Simulation Parameters}
    \label{tableSimulationParameters}
    \begin{tabular}{l l}
        \toprule
        \textbf{Parameter} & \textbf{Value} \\
        \midrule
        Total number of antenna elements, $M$ & 16 - 96\\
        Number of antenna elements on each AP, $S$ & 4\\
        Number of APs, $Q$ & 4 - 16\\
        Number of active MTDs, $K$ & 16 - 64\\
        Length of the side of the square area, $l$ & 250 m - 1 km\\
        Height of the BS or AP, $h$ & 6 m\\
        Height of the MTDs, $h_{\text{MTD}}$ & 1.5 m\\
        Target data rate, $R$ & 1 bit/s/Hz \\
        Carrier frequency, $f_c$ & 3.5 GHz\\
        Transmit power of the MTDs, $p_u$ & 20 dBm\\
        PSD of the thermal noise, $N_0$ & -174 dBm/Hz\\        
        Bandwidth, $B$ & 20 MHz\\
        Noise figure at the receivers, $N_F$ & 7 dB\\
        Number of simultaneous alarm events, $A$ & 1\\
        Epicenter of the alarm event, $(x_{a},y_{a},z_{a})$ & $l/4,l/4,0$\\
        Intensity of the alarm event, $\nu$ & 50\\
        \bottomrule
    \end{tabular}    
\end{table}

\par
\begin{figure}
    \centering
    \includegraphics[scale=0.4]{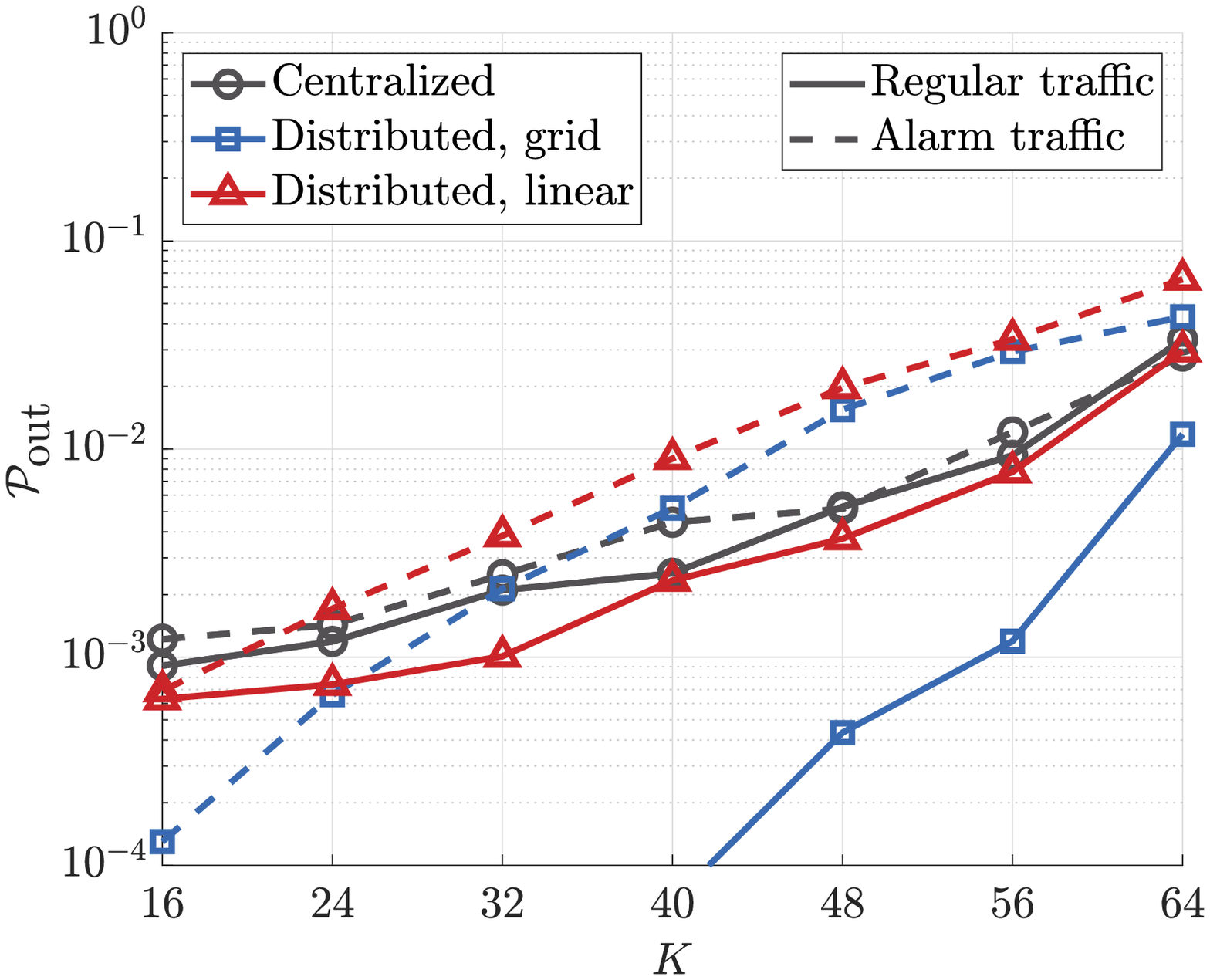}
    \caption{Outage probability versus number of active MTDs for $M=64$, $l=250$ m and centralized MMSE combining.}
    \label{Results_Massive_Connectivity}
    \vspace{0.5cm}

    \includegraphics[scale=0.4]{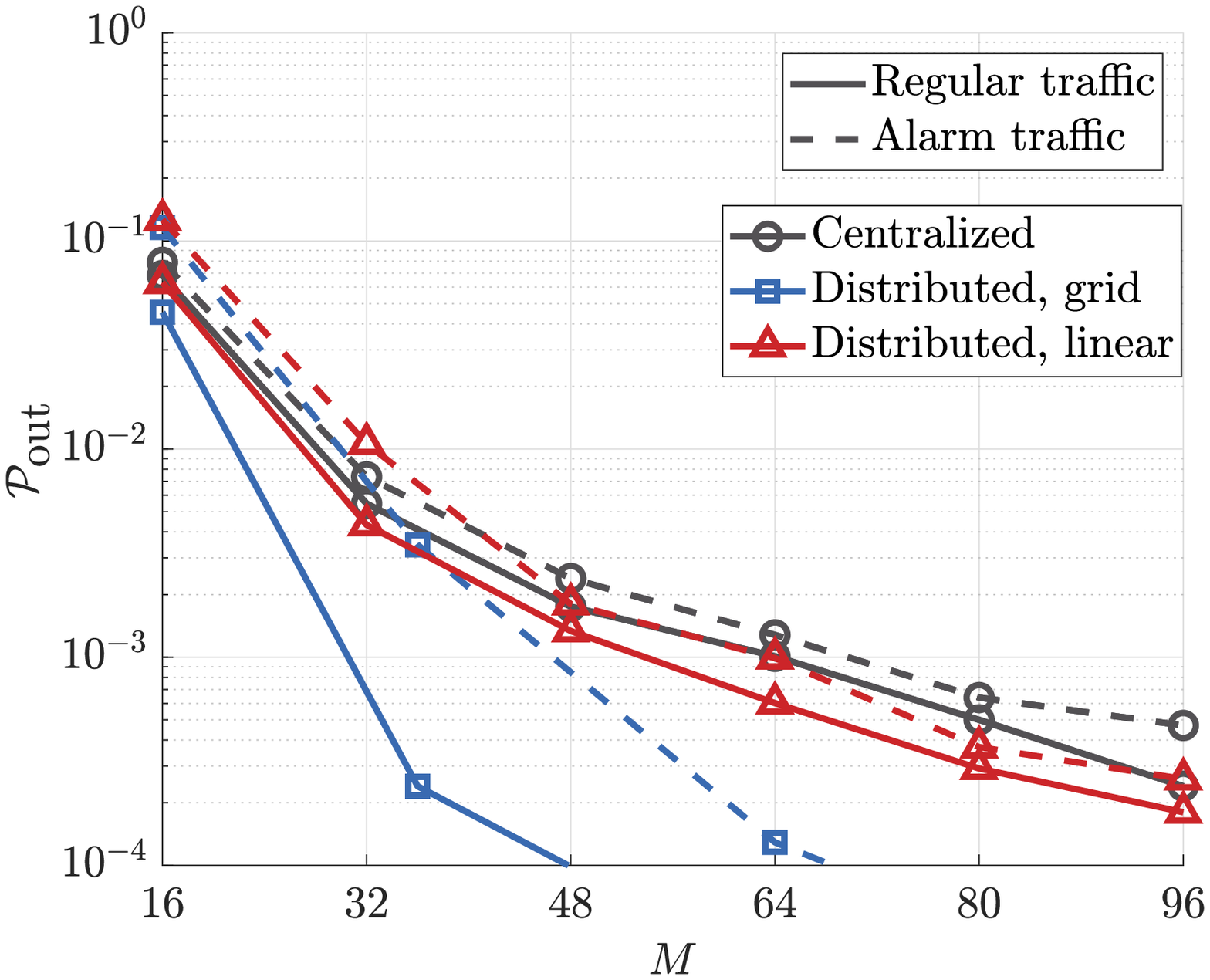}
    \caption{Outage probability versus number of antenna elements for $K=16$, $l=250$ m and centralized MMSE combining.}
    \label{Results_Ultra_Reliability}
    \vspace{0.5cm}

    \includegraphics[scale=0.4]{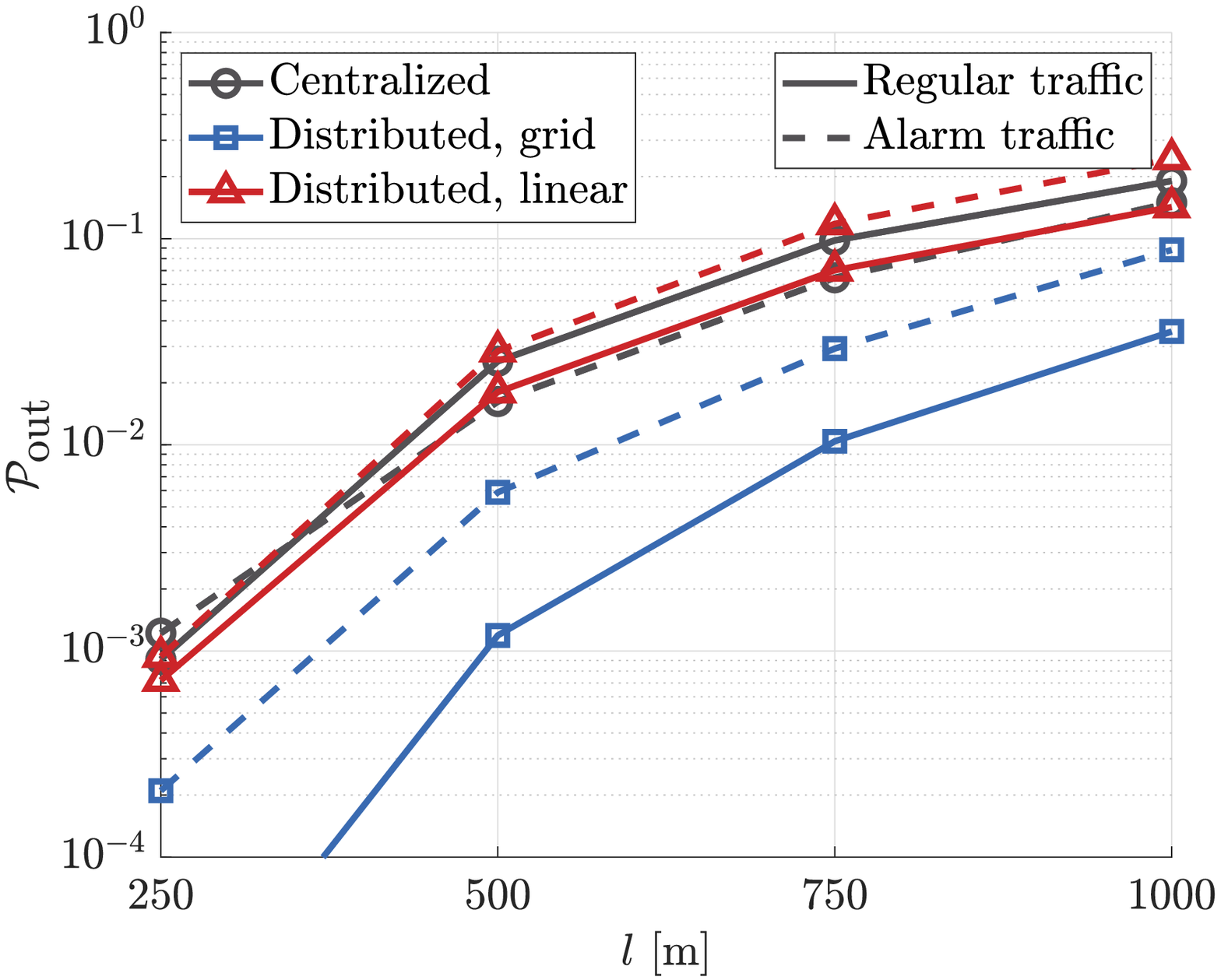}
    \caption{Outage probability versus dimensions of the site for $M=64$ $K=16$, and centralized MMSE combining.}
    \label{Results_Sustainable_Coverage}
\end{figure}

 Fig. \ref{Results_Massive_Connectivity} shows the outage probability versus $K$ for $M=64$ and $l=250$ m, Fig. \ref{Results_Ultra_Reliability} shows the outage probability versus $M$ for $K=16$ and $l=250$ m, and Fig. \ref{Results_Sustainable_Coverage} shows the outage probability versus $l$ for $M=64$ and $K=16$. From the curves related to regular traffic, we obtain the same well-known findings from the literature: CmMIMO is always outperformed by DmMIMO, and DmMIMO with the grid configuration of APs presents better performance than the linear deployment owing to the higher macro-diversity gains. However, even though the grid deployment outperforms the linear deployment, it requires higher fronthaul capacity. In a linear deployment, only a single fronthaul connection to the CPU is required, while each AP needs to have its own connection to the CPU in a grid deployment.

\par From the curves related to alarm traffic, we obtain some interesting findings. CmMIMO can outperform DmMIMO for some combinations of the parameters $[M,K,l]$. When an alarm event occurs, most of the active MTDs are close to the epicenter of the event. As a consequence, in a DmMIMO network, most of the antenna elements are far away from the epicenter, thus providing poor coverage to the active MTDs. In this case, CmMIMO can outperform DmMIMO because the position of the BS in the center of the square area decreases the average distance between the antenna elements and the MTDs. Thus, as shown in Fig. \ref{Results_Ultra_Reliability}, DmMIMO can only outperform CmMIMO under alarm traffic if the total number of antenna elements is high enough to guarantee that many antenna elements will be close to the position of the alarm event.

\section{Conclusions}
\label{Conclusions}


\par In this work, we compared the performance of CmMIMO and DmMIMO in an indoor industrial scenario. We considered two different spatial distributions of APs for DmMIMO: a grid configuration of APs and a linear deployment. We also studied the performance of CmMIMO and DmMIMO under distinct traffic models for MTC: regular traffic and alarm traffic. Our simulation results showed that under regular traffic, DmMIMO deployments consistently outperform CmMIMO. DmMIMO with grid deployment outperforms the linear deployment, although requiring a higher fronthaul capacity. Under alarm traffic, most active devices are very close to the epicenter of an alarm event. Owing to the spatial distribution of APs, most of the antenna elements in a DmMIMO network are far away from the epicenter of the alarm event, thus CmMIMO can outperform DmMIMO in this situation.


\section*{Acknowledgment}

\par We thank the Academy of Finland, 6Genesis Flagship (grant no. 346208), European Union’s Horizon 2020 research and innovation programme (EU-H2020), Hexa-X (grant no. 101015956) and Hexa-X-II (grant no. 101095759) projects, the Finnish Foundation for Technology Promotion, and CNPq (Brazil).

\bibliographystyle{./bibliography/IEEEtran}
\bibliography{./bibliography/references}

\end{document}